\documentclass[aps,prd,twocolumn,10pt,altaffilletter,superscriptaddress,showpacs,nofootinbib,preprintnumbers]{revtex4-1}
\pdfoutput=1

\usepackage{graphicx,color,amsfonts,amssymb,lipsum,footnote,bm,keyval,hyperref,tikz,amsmath}
\hypersetup{pdfpagemode=UseNone}

\newcommand{\be}{\begin{equation}} 
\newcommand{\ee}{\end{equation}} 
\newcommand{\bea}{\begin{eqnarray}} 
\newcommand{\eea}{\end{eqnarray}} 
\newcommand{\nn}{\nonumber} 
\newcommand{\mintedim}[2]{{\int\kern-0.50em\mbox{{\small$\mathop{\frac{\mbox{{\small${\rm d^{#2}}\vect{#1}$}}}{\mbox{{\small$(2\pi)^{#2}$}}}}$}}\ }} 
\newcommand{\inteonedim}[1]{{\int_0^\infty\kern-1em\mbox{{\small${\rm d}{#1}$}}}} 
\newcommand{\intecontour}[2]{{\int_{{#1}-i\infty}^{{#1}+i\infty}{\kern-2.25em d{#2}/(2i\pi)}}} 
\newcommand{\vect}[1]{\bm{#1}} 

\begin{document}

\title{Non-extensive Fokker-Planck transport coefficients of heavy quarks }

\author{Trambak~Bhattacharyya}
\email{trambak.bhattacharyya@uct.ac.za}
\affiliation{UCT-CERN Research Centre, University of Cape Town, Rondebosch 7701, South Africa}
\affiliation{Department of Physics, University of Cape Town, Rondebosch 7701, South Africa}

\author{Jean~Cleymans}
\email{jean.cleymans@uct.ac.za}
\affiliation{UCT-CERN Research Centre, University of Cape Town, Rondebosch 7701, South Africa}
\affiliation{Department of Physics, University of Cape Town, Rondebosch 7701, South Africa}

\everymath{\displaystyle}

\begin{abstract}
In presence of the non-ideal plasma effects, Heavy Quarks (HQs) carry out non linear random walk inside Quark-Gluon Plasma (QGP) and in the small momentum transfer limit, the evolution of the HQ distribution is dictated by the Non Linear Fokker-Planck Equation (NLFPE). Using the NLFPE, we calculate the transport coefficients (drag and diffusion) of heavy quarks travelling through QGP. We observe substantial modification in the momentum and temperature variation of the transport coefficients; and this will modify the physical picture we are having about the transport of heavy quarks inside QGP, and hence, about the characterisation of the plasma.
\end{abstract}

\maketitle


\section{Introduction}
\label{intro}
Nuclear matter at sufficiently high temperature and energy density exists in a form where the quarks and gluons are not confined. This form of matter created in RHIC at BNL and LHC at CERN, Geneva is called Quark Gluon Plasma (QGP). Characterization of the medium is a very vigorous field of research nowadays and different methods are utilised in doing so. One of them is to study the energy loss of high energy probes while travelling through the QGP. Highly energetic heavy quarks (charm and bottom) act as very clean probes in this respect because they are not produced in plentitude and hence the bulk properties of the medium are not governed by them. Also, they are produced very early in the collision to witness the entire evolution. Apart from that, heavy quarks barely thermalize with the medium particles and hence, it is relatively easy to observe the change of their distribution which can be associated with the experimental observations like nuclear modification factor ($R_{\mathrm{AA}}$). 

The evolution of the heavy quark distribution inside QGP is dictated by the Boltzmann Transport Equation (BTE). Heavy quarks inside QGP can interact via two primary processes i) collisional interaction and ii) radiative interaction. The importance of the collisional energy loss in characterizing QGP was first pointed by Bjorken \cite{bjorken}. While improved estimations of the collisional energy loss were done in Refs. \cite{thomagyulassy}-\cite{mgmthoma}, the radiative energy loss has been the subject matter of Refs. \cite{mgmplbeloss}-\cite{pshuklaeloss}.
Assuming small momentum transfer between the probe and the medium, BTE can be reduced to the Fokker-Planck Equation (FPE). The evolution is caused by the interaction of the heavy quarks with the medium particles and they are encoded in the transport coefficients like drag and diffusion which act as the inputs to the Fokker-Planck equation. Energy loss is related to the drag and diffusion coefficients which are again related through the fluctuation-dissipation theorem \cite{rafelskiwaltonprl}. Hence, precise evaluation of drag and diffusion coefficients of heavy quarks  are necessary for the characterization of plasma.

The problem of evaluating the Fokker-Planck drag and diffusion coefficients of heavy flavours has been treated in a number of research papers \cite{chaksyamnuovocimento}-\cite{bkpatraepja}. All of them use the linear Fokker-Planck equation which is derived from the linear Boltzmann transport equation assuming small momentum transfer among the probe and the medium particles. It is believed that only the exponential distribution can be the stationary solution of the BTE; but in presence of non-ideal plasma effects (e.g long range correlations) BTE can be modified and the stationary solution of the modified BTE is the Tsallis like (power law) distribution proposed in \cite{tsallis}.

One of the assumptions which makes the conventional BTE (which, henceforth, will be called as linear BTE or LBTE) to have exponential stationary solution is the `molecular chaos hypothesis'. This hypothesis says that the two particle distribution (appearing in the collision term of the BTE) can be factorized into two one particle distributions, {\it i.e} those of the probe and the medium particle \cite{liboffbook}. The collision term in the modified  (or non-linear) BTE, to be called as NLBTE, generalizes this hypothesis. Many recent references have treated this modified form of the Boltzmann transport equation in different contexts \cite{lavagnopla}-  \cite{Biro:2012ix}. In this paper, we treat the transport of heavy quarks inside quark gluon plasma with the help of the Non Linear Fokker-Planck Equation (NLFPE) which is obtained as the small momentum transfer limit approximation of the NLBTE. The non linear Fokker Planck drag and diffusion coefficents will be evaluated.

The article is organized as follows: in the next section we set up the formalism for computing the Non Linear Fokker Planck Transport Coefficients (NLFPTC). In Sec. \ref{results} we show the plots depicting the variation of the NLFPTC with the incoming (probe) particle momentum and with the medium temperature. In Sec. \ref{summary}, we summarize and indicate possible future directions.


\section{Formalism}
\label{formalism}

\subsection{Non Linear Boltzmann Transport Equation}
\label{nlbte}

To begin with, we will outline the derivation of NLBTE following \cite{lavagnopla}. Within the Tsallis statistical mechanics framework (also called the Non Extensive (NE) framework), the particle four flow can be defined as:

\be
N^{\mathrm{NE}}_{\mu}= c \int \frac{d^3p}{p^0} p_{\mu} \tilde{f}^{\mathrm{NE}}(x,p)
\label{particlefourflow}
\ee
where ${f}^{\mathrm{NE}}$ is the non extensive version of the phase space distribution of the probe particle with four momentum $p$ at the space time point $x$; and $\tilde{f}^{\mathrm{NE}}=\left(f^{\mathrm{NE}}\right)^{q_{\mathrm{T}}}$, where $q_{\mathrm{T}}$ is the Tsallis parameter. The 0-th component of $N^{\mathrm{NE}}_{\mu}$ gives the thermodynamically consistent particle number density per degree of freedom (times a constant $c>0$) in the Tsallis statistics \cite{cleymansworkuepja}. From Eq. \eqref{particlefourflow}, we can form a scalar quantity

\be
\Delta N^{\mathrm{NE}}= \int_{\Delta^3\sigma}  \int_{\Delta^3 p}  d^3\sigma_{\mu }\frac{d^3p}{p^0} p^{\mu} \tilde{f}^{\mathrm{NE}}(x,p)
\label{particlefourflow}
\ee
where $d^3\sigma_{\mu}$ is the element of a time like three surface and $\Delta^3\sigma$ is a small element situated at $x$. We can explain 
$\Delta N^{\mathrm{NE}}$ as the net flow passing through a segment of $\Delta^3\sigma$ with momentum range ${\Delta^3 p} $ around $p$. 
Considering collisions among the particles, the net flow through $\Delta^3\sigma$ can be written as \cite{degroot}:

\bea
p^{\mu}\partial_{\mu}\tilde{f}^{\mathrm{NE}}(x,p)&=& 
\left(\frac{\partial}{\partial t} +\frac{{\bf p}}{E_{\bf p}}.\frac{\partial}{\partial {\bf x}}+{\bf F}.\frac{\partial}{\partial {\bf p}}\right) 
\tilde{f}^{\mathrm{NE}}(x,p)\nn\\
&=& C^{\mathrm{NE}}
\label{nebte}
\eea
where $E_{\bf p}$ is the energy of the incoming particle (heavy quark) and $C^{\mathrm{NE}}$ is the non extensive collision term. Eq. \eqref{nebte} is the desired kinetic equation and is called the non linear Boltzmann transport equation .

Assuming that the change in the distribution function is due to the binary collisions only, the non extensive collision term $C^{\mathrm{NE}}$
in the NEBTE is given by \cite{lavagnopla}-  \cite{Biro:2012ix}:

\bea
C^{\mathrm{NE}}
&=& \frac{1}{2E_p} \int \frac{d^3\bf{q}}{(2\pi)^3} \frac{d^3{\bf q'}} {(2\pi)^3} \frac{d^3{\bf p'}} {(2\pi)^3}  |\overline{M}|^2 (2\pi)^4 \nn\\
&&\delta^4(p+q-p^{'}-q^{'}) \times \nn\\
&& \left[h_{\mathrm{NE}} (f^{\mathrm{NE}}(x,p'),f^{\mathrm{NE}}(x,q')) \right. \nn\\
&&\left. -h_{\mathrm{NE}} (f^{\mathrm{NE}}(x,p),f^{\mathrm{NE}}(x,q)) \right]
\label{necollterm}
\eea
where $p(q)$ is the incoming four momentum and $p'(q')$ is the outgoing four momentum. The quantity $h_{\mathrm{NE}}  (f^{\mathrm{NE}}(x,p'),f^{\mathrm{NE}}(x,q') )$ represents the two particle distribution function  \cite{liboffbook} with four momenta $p'$ and $q'$ at the same space time point $x\equiv({\bf x},t)$. The function $h_{\mathrm{NE}}$ can be defined in the following way:

\bea
h_{\mathrm{NE}}(f_a,f_b) = \mathrm{Exp}_{\mathrm{NE}} \left[ \mathrm{log}_{\mathrm{NE}} (f_a)
+ \mathrm{log}_{\mathrm{NE}}(f_b)\right]
\label{defh}
\eea

In the conventional Boltzmann transport equation $h_{\mathrm{NE}}$ is replaced by the product of the two distribution functions and this replacement can be done under the assumption of the `molecular chaos'. To find how the modified collision term looks like, we define the following quantities: 

\begin{itemize}

\item the three momentum transfer $\bf{k}=\bf{p}-\bf{p}^{'}=\bf{q}^{'}-\bf{q}$, which is the spatial part of the four momentum transfer $k=p-p'=q'-q$ ;

\item the non extensive exponential as well as the non extensive logarithm function:

\bea
\mathrm{Exp}_{\mathrm{NE}}(x)=(1- \delta q ~x)^{-\frac{1}{\delta q}} ~&;&~\mathrm{log}_{\mathrm{NE}}(x) = \frac{1-x^{-\delta q}}{\delta q}\nn\\&& 
\label{defneexplog} 
\eea
where $\delta q=q_{\mathrm{T}}-1>0$ 
\end{itemize}

Using the definitions in Eq. \eqref{defneexplog}, Eq. \eqref{defh} can be expanded in a series of $\delta q$:

\bea
h_{\mathrm{NE}}(f_a,f_b) &=& \left\{1 - \delta q \left(  \frac{1-f_a^{-\delta q}}{\delta q}  +  \frac{1-f_b^{-\delta q}}{\delta q}     \right) \right\}^{-\frac{1}{\delta q}} \nn\\
&=& f_a f_b+\delta q ~f_a f_b \log \left(f_a\right) \log
   \left(f_b\right)+O\left({\delta q}^2\right) \nn\\
\eea
whose first term gives back the original Boltzmann transport equation obtained using the `molecular chaos hypothesis'. Also, we note that the Tsallis non extensive parameter $\delta q$ acts as the correlation between the two distribution functions. Hence, $h_{\mathrm{NE}}(f_a,f_b)$ can be thought of as the two-particle distribution which can be expressed as the product of two single particle distributions added to the correlation; and of course

\bea
\lim_{\delta q\rightarrow 0}~~h_{\mathrm{NE}}(f_a,f_b)=f_a f_b
\eea

Defining,

\bea
\hat{\Phi} [..]&=& \frac{1}{2E_p} \int \frac{d^3{\bf q}}{(2\pi)^3} \frac{d^3{\bf q'}} {(2\pi)^3} \frac{d^3{\bf p'}} {(2\pi)^3}  |\overline{M}|^2 (2\pi)^4 \nn\\
&&\times \delta^4(p+q-p^{'}-q^{'}) [..]
\eea
and
\bea
\mathcal{D}^{\mathrm{NE}}=\left[h_{\mathrm{NE}} (f^{\mathrm{NE}}(x,p'),f^{\mathrm{NE}}(x,q')) \right. \nn\\
\left. -h_{\mathrm{NE}} (f^{\mathrm{NE}}(x,p),f^{\mathrm{NE}}(x,q)) \right]
\eea
the collision term $C^{\mathrm{NE}}$ can be written as:

\bea
C^{\mathrm{NE}} &=& \frac{1}{2E_p} \int \frac{d^3\bf{q}}{(2\pi)^3} \frac{d^3{\bf q'}} {(2\pi)^3} \frac{d^3{\bf p'}} {(2\pi)^3}  |\overline{M}|^2 \nn\\
&&\times (2\pi)^4 \delta^4(p+q-p^{'}-q^{'}) \times \nn\\
&&  \mathcal{D}^{\mathrm{NE}}
\nn\\
&=& \hat{\Phi} \mathcal{D}^{\mathrm{NE}}
\label{nonextcollterm}
\eea


\subsection{Space Averaging}
\label{spav}

At this stage, we would like to average the distribution functions over the space and will work with the momentum space distribution function only. While doing so, we will assume that the phase space distribution of the heavy quarks (given by $f^{\mathrm{NE}}(x,p)$ or $f^{\mathrm{NE}}(x,p')$) can be inhomogeneous but that of the medium particles (given by $f^{\mathrm{NE}}(x,q)$ or $f^{\mathrm{NE}}(x,q')$) is independent of $x$. Also, we assume that the external force ${\bf F}=0$. After space averaging and putting ${\bf v}={\bf p}/E_{\bf p}$, Eq. \eqref{nebte} can be written as:

\begin{widetext}
\bea
\frac{1}{V} &&\int d^3{\bf x} \left(\frac{\partial}{\partial t} \right.+ \left. {\bf v}.\frac{\partial}{\partial {\bf x}}\right) \tilde{f}^{\mathrm{NE}} =
\frac{1}{V} \int d^3{\bf x} ~C^{\mathrm{NE}} ~~~~
~~~\Rightarrow  \frac{\partial}{\partial t} f_{\bf p}^{\mathrm{}NE} + \int d^3{\bf x} ~{\bf v}.\frac{\partial}{\partial {\bf x}} \tilde{f}^{\mathrm{NE}} =
\hat{\Phi}  \int d^3{\bf x} ~ \mathcal{D}^{\mathrm{NE}} \nn\\
\Rightarrow  \frac{\partial}{\partial t} f_{\bf p}^{\mathrm{NE}} &+& \int_{-\infty}^{\infty} dx~ dy~ dz~ \left( v_x  \frac{\partial}{\partial x}  \tilde{f}^{\mathrm{NE}} + v_y  \frac{\partial}{\partial y}  \tilde{f}^{\mathrm{NE}} + v_z  \frac{\partial}{\partial z}  \tilde{f}^{\mathrm{NE}}\right) 
 = \hat{\Phi}  \mathcal{D}^{\mathrm{NE}}_{\bf p} \nn\\
 \Rightarrow 
  \frac{\partial}{\partial t} f_{\bf p}^{\mathrm{NE}} &+& \left[ \int_{-\infty}^{\infty} dy~ dz~ v_x \left\{f(\infty,y,z,t,{\bf p})-f(-\infty,y,z,t,{\bf p})\right\} 
  +\int_{-\infty}^{\infty} dx~ dz~ v_y \left\{f(x,\infty,z,t,{\bf p})-f(x,-\infty,z,t,{\bf p})\right\} \right. \nn\\ 
&&  \left. +\int_{-\infty}^{\infty} dx~ dy~ v_z \left\{f(x,y,\infty,t,{\bf p})-f(x,y,-\infty,t,{\bf p})\right\}\right] =\hat{\Phi}  \mathcal{D}^{\mathrm{NE}}_{\bf p} \nn\\
\eea
\end{widetext}
assuming that $v_x,~v_y,~v_z$ are not the functions of ${\bf x}$. Now, the distribution function is symmetric in ${\bf x}$  and it vanishes at $x,~y,~z=\pm\infty$ \footnote{These are determined by the temperature profile we choose here.}; and so we are left with the following equation.

\bea
 \frac{\partial}{\partial t} f_{\bf p}^{\mathrm{NE}} = \hat{\Phi}  \mathcal{D}^{\mathrm{NE}}_{\bf p} 
 \label{nebtespav}
\eea
 where $f_{\bf p}^{\mathrm{NE}}$ and  $\mathcal{D}^{\mathrm{NE}}_{\bf p} $ are the space averages of the functions $\tilde{f}^{\mathrm{NE}}$ and 
 $\mathcal{D}^{\mathrm{NE}}$.
Hence, with the help of the Eq. \eqref{nebtespav} we can study the evolution of the momentum distribution of the heavy quarks produced due to very early hard processes and carrying out the random motion in a medium of light quarks and gluons. Remembering the fact that the genesis of the heavy quarks is due to the hard processes, their momentum distribution can be characterised by a power law distribution, the Tsallis distribution in the present case \cite{tsallisraaepja}. Now, we know that the Landau Kinetic Approximation (LKA) (${\bf k} \rightarrow 0$, which essentially means that the step size of the heavy quark carrying out the momentum space random motion in the medium is vanishingly small \cite{nefpe_curado}) of the collision term in the extensive BTE gives rise to the linear Fokker-Planck Transport Coefficients \cite{svetitsky} like drag and diffusion. A similar approach can be taken to derive the expressions for the NLFPTC (or non extensive Fokker-Planck transport coefficients) using Eqs. (\ref{defh},  \ref{defneexplog}, \ref{nonextcollterm}). 

The space averaging of the non extensive collision term involves the knowledge of the spatial and the temporal variation of temperature which essentially gives rise to the spatial and the temporal variation of the collision term. Here we use the temperature profile used in \cite{tempevolution}.  

\bea
T({\bf x};t)=\frac{T_{p}(t)}{\left[1+\mathrm{Exp}\left\{a(t)\left(\frac{\sqrt{x^2+y^2+z^2}}{r_0(t)}-1\right)\right\}\right]}
\label{tempprof}
\eea
for the parameters $T_p,~a~\mathrm{and}~r_0>0$.  The temperature profile makes the phase space distribution vanish at infinity. 
Hence, with the help of the temperature profile in Eq. \eqref{tempprof}, we are now ready to perform the space averaging ritual. For the extensive case, the distribution of the probe particle and that of the medium come as a product and hence, the space averaging is relatively simpler to perform. But now we have a non-trivial interplay between the two distribution functions; and so, we detail it in the following few paragraphs.

The result of the space averages of $\tilde{f}^{\mathrm{NE}}$ and $\mathcal{D}^{\mathrm{NE}}$ in the non extensive case is obtained in terms of the hypergeometric function `$_4F_3$'.  Defining the following quantities:

\bea
g_{1,{\bf p}} &=& 1+ \frac{E_{\bf p}\delta q}{T_p} \\
g_{2,{\bf p}} &=& \frac{\delta q}{T_p} \mathrm{Exp}(-a)E_{\bf p} \\
g_{3,{\bf p,q}} &=& 1+ \frac{E_{\bf p}\delta q}{T_p}+\frac{E_{\bf q}\delta q}{T_q}\\
c&=&\frac{1+\delta q}{\delta q}
\eea
we will express the space averages in terms of them.

First of all, we can write down the space average of $\tilde{f}^{\mathrm{NE}}$ in the following form  (for chemical potential $\mu$=0):

\bea
f_{\bf p}^{\mathrm{NE}} &\approx& \int d^3{\bf x} \left(1+\frac{\delta q}{T({\bf x};t)} E_{\bf p}\right)^{-\frac{1+\delta q}{\delta q}}
\label{fp}\nn\\
&=& g_{2,{\bf p}}^{-c}\frac{r_0^3}{a^3c^3}  ~_4F_3\left[c,c,c,c;c+1,c+1,c+1;- \frac{g_{1,{\bf p}}}{g_{2,{\bf p}}}\right] \nn\\
\eea

After getting the space averaged momentum distribution of the heavy quarks we will perform the space averaging of the collision term; and while doing so we replace $p'=p-k;~q'=q+k$ 

\begin{widetext}
\begin{eqnarray}
\hat{\Phi} \mathcal{D}_{\bf p}^{\mathrm{NE}} &=& \int d^3{\bf x} ~\hat{\Phi} \left[h_{\mathrm{NE}} (f^{\mathrm{NE}}(x,p-k),f^{\mathrm{NE}}(x,q+k))-h_{\mathrm{NE}} (f^{\mathrm{NE}}(x,p),f^{\mathrm{NE}}(x,q)) \right] \nn\\
&\approx&  \int d^3{\bf x} ~ \hat{\Phi}  \left[ \mathrm{Exp}_{\mathrm{NE}} \left\{ \mathrm{log}_{\mathrm{NE}} \left( 1+\frac{\delta q}{T({\bf x};t)} E_{\bf p-k} \right)^{-\frac{1}{\delta q}} +\mathrm{log}_{\mathrm{NE}} \left( 1+\frac{\delta q}{T_{\mathrm{q}}} E_{{\bf q+k}} \right)^{-\frac{1}{\delta q}}     \right\} \right. \nn\\
&& \left. -  \mathrm{Exp}_{\mathrm{NE}} \left\{  \mathrm{log}_{\mathrm{NE}} \left( 1+\frac{\delta q}{T({\bf x};t)} E_{{\bf p}} \right)^{-\frac{1}{\delta q}} +\mathrm{log}_{\mathrm{NE}} \left( 1+\frac{\delta q}{T_{\mathrm{q}}} E_{{\bf q}} \right)^{-\frac{1}{\delta q}}  \right\} \right] \nn\\
&=& \int d^3{\bf x} ~  \hat{\Phi}  \left[ \mathrm{Exp}_{\mathrm{NE}} \left( -\frac{ E_{\bf p-k}}{T({\bf x};t)} - \frac{ E_{\bf q+k}}{T_{\mathrm{q}}} \right) -  
\mathrm{Exp}_{\mathrm{NE}} \left( -\frac{ E_{\bf p}}{T({\bf x};t)} - \frac{ E_{\bf q}}{T_{\mathrm{q}}}\right) \right] \nn\\
&=& \hat{\Phi} \left\{g_{2,{\bf p-k}}^{-c+1}\frac{r_0^3}{a^3(c-1)^3}  ~_4F_3\left[c-1,c-1,c-1,c-1;c,c,c;- \frac{g_{3,{\bf p-k,q+k}}}{g_{2,{\bf p,q}}}\right] \right.\nn\\
&&\left.- g_{2,{\bf p}}^{-c+1}\frac{r_0^3}{a^3(c-1)^3}  ~_4F_3\left[c-1,c-1,c-1,c-1;c,c,c;- \frac{g_{3,{\bf p,q}}}{g_{2,{\bf p}}}\right] \right\} \nn\\
&=& \hat{\Phi} \left[f_{\bf p-k}^{\mathrm{NE}}\mathcal{R}^1_{\bf p-k,q+k}-f_{\bf p}^{\mathrm{NE}}\mathcal{R}^1_{\bf p,q}\right]\label{colltermreduced}\\
\mathrm{For,~}\mathcal{R}^1_{\bf p,q}&=& \left( f_{\bf p}^{\mathrm{NE}}\right)^{-1} g_{2,{\bf p}}^{-c+1}\frac{r_0^3}{a^3(c-1)^3}  ~_4F_3\left[c-1,c-1,c-1,c-1;c,c,c;- \frac{g_{3,{\bf p,q}}}{g_{2,{\bf p}}}\right] \nn
\label{colltermspav}
\end{eqnarray}
\end{widetext}

 The approximate equalities in Eqs. (\ref{fp}, \ref{colltermspav}) come when we assume that the Tsallis parameters characterizing the momentum distributions of the probe particle (momenta $p$ and $p'$) and the medium particle (momenta $q$ and $q'$) are almost same with $\delta q$ which characterizes the correlation between them. This special case gives rise to the close analytical results upon space averaging in terms of the hypergeometric functions; but no closed form exists when we are off from this assumption.


\subsection{The Non Linear Fokker Planck Transport Coefficients}
\label{nlfptc}

Expanding the r.h.s. of the Eq. (\ref{colltermreduced}) in the Taylor's expansion for small ${\bf k}$ we get,

\begin{eqnarray}
\hat{\Phi} \mathcal{D}^{\mathrm{NE}} &\approx&  \hat{\Phi}  f^{\mathrm{NE}}_{\bf p} \mathcal{R}^1_{\bf p,q} +\frac{\partial}{\partial p_i} \left[ \hat{\Phi} \{-{\bf k_i}\} f^{\mathrm{NE}}_{\bf p} \mathcal{R}^1_{\bf p,q} \right] + \nn\\
&&\frac{\partial^2}{\partial p_i \partial p_j} \left[  \hat{\Phi} \left\{\frac{1}{2}  {\bf k}_i {\bf k}_j \right\}  \left( f^{\mathrm{NE}}_{\bf p} \right)^{1-\delta q}  
\mathcal{R}^2_{\bf p,q}\right] -  \nn\\ && \hat{\Phi}  f^{\mathrm{NE}}_{\bf p} \mathcal{F}_{\bf p,~q}  \nn\\
&=& \frac{\partial}{\partial p_i} \left[ \hat{\Phi} \{-{\bf k_i}\} f^{\mathrm{NE}}_{\bf p} \mathcal{R}^1_{\bf p,q} \right] \nn\\
&+&  \frac{\partial^2}{\partial p_i \partial p_j} \left[  \hat{\Phi} \left\{\frac{1}{2}  {\bf k}_i {\bf k}_j \right\}  \left( f^{\mathrm{NE}}_{\bf p} \right)^{1-\delta q} \mathcal{R}^2_{\bf p,q} \right] \label{nefpetaylors}\\
\mathrm{For,~} \mathcal{R}^2_{\bf p,q} &=& \left( f^{\mathrm{NE}}_{\bf p} \right)^{\delta q} \mathcal{R}^1_{\bf p,q} \nn
\end{eqnarray}

The r.h.s. of Eq. (\ref{nefpetaylors}) can be equated with the Non Extensive Fokker Planck Equation (NEFPE) in the momentum space which is given by

\begin{eqnarray}
\frac{\partial f^{\mathrm{NE}}_{\bf p}}{\partial t}  = -\frac{\partial}{\partial p_i} \left[ A_i^{\mathrm{NE}} f^{\mathrm{NE}}_{\bf p}  \right] + \frac{\partial^2}{\partial p_i \partial p_j} \left[ B_{ij}^{\mathrm{NE}} \left(f^{\mathrm{NE}}_{\bf p} \right)^{1-\delta q} \right] \nn\\
\label{nefpeq}
\end{eqnarray}
This form of NEFPE has been used in the Refs. \cite{wolschinplb}, \cite{lavagnobrazjour}.
Comparing the single and double derivative terms in the Eqs. (\ref{nefpetaylors}, \ref{nefpeq}), we get the following expressions for the non-extensive Fokker-Planck drag ($A$) and diffusion coefficients ($B$):

\begin{eqnarray}
A_i^{\mathrm{NE}} &=& \hat{\Phi} \left[{\bf k_i} ~ \mathcal{R}^1_{\bf p,q}\right] \nn\\
B_{ij}^{\mathrm{NE}} &=& \hat{\Phi} \left[\frac{1}{2}{\bf k_i~k_j}  ~\mathcal{R}^2_{\bf p,q} \right] 
\label{abshort}
\end{eqnarray}

The longhand expressions for the above two non extensive quantities are given below. Also, for comparison, we tabulate their extensive counterparts \cite{svetitsky}, too.

\begin{widetext}
\subsubsection{Non Extensive}
\bea
A_i^{\mathrm{NE}} &=&  \frac{1}{2E_{\bf p}} \int \frac{d^3{\bf q}}{(2\pi)^3} \frac{d^3{\bf q'}} {(2\pi)^3} \frac{d^3{\bf p'}} {(2\pi)^3}  |\overline{M}|^2 (2\pi)^4 \delta^4(p+q-p^{'}-q^{'}) \nn\\
&&\times \underbrace{\left( f_{\bf p}^{\mathrm{NE}}\right)^{-1} g_{2,{\bf p}}^{-c+1}\frac{r_0^3}{a^3(c-1)^3}  ~_4F_3\left[c-1,c-1,c-1,c-1;c,c,c;- \frac{g_{3,{\bf p,q}}}{g_{2,{\bf p}}}\right]}_{\mathcal{R}^1_{\bf p,q}}    ({\bf p-p'})_i \nn\\
B_{ij}^{\mathrm{NE}} &=&  \frac{1}{2E_{\bf p}} \int \frac{d^3{\bf q}}{(2\pi)^3} \frac{d^3{\bf q'}} {(2\pi)^3} \frac{d^3{\bf p'}} {(2\pi)^3}  |\overline{M}|^2 (2\pi)^4 \delta^4(p+q-p^{'}-q^{'}) \nn\\
&& \underbrace{\left( f_{\bf p}^{\mathrm{NE}}\right)^{-1+\delta q} g_{2,{\bf p}}^{-c+1}\frac{r_0^3}{a^3(c-1)^3}  ~_4F_3\left[c-1,c-1,c-1,c-1;c,c,c;- \frac{g_{3,{\bf p,q}}}{g_{2,{\bf p}}}\right]}_{\mathcal{R}^2_{\bf p,q}}  \frac{1}{2}({\bf p'-p})_i ({\bf p'-p})_j 
\label{nonexdragdiff}
\eea

\subsubsection{Extensive}

\bea
A_i&=& \frac{1}{2E_{\bf p}} \int \frac{d^3{\bf q}}{(2\pi)^3} \frac{d^3{\bf q'}} {(2\pi)^3} \frac{d^3{\bf p'}} {(2\pi)^3}  |\overline{M}|^2 (2\pi)^4 \delta^4(p+q-p^{'}-q^{'})~~ f({\bf q})~~ ({\bf p-p'})_i \nn\\
B_{ij}&=&  \frac{1}{2E_{\bf p}} \int \frac{d^3{\bf q}}{(2\pi)^3} \frac{d^3{\bf q'}} {(2\pi)^3} \frac{d^3{\bf p'}} {(2\pi)^3}  |\overline{M}|^2 (2\pi)^4 \delta^4(p+q-p^{'}-q^{'}) ~~f({\bf q})~~ \frac{1}{2}({\bf p'-p})_i ({\bf p'-p})_j 
\label{exdragdiff}
\eea
\end{widetext}

$A_i^{\mathrm{NE}}$ and $B_{ij}^{\mathrm{NE}}$ depend only on the vector ${\bf p}$ and we can write them as a combination of the Kronecker's delta $\delta_{ij}$ and ${\bf p}$:

\bea
A_i^{\mathrm{NE}} &=& {\bf p_i} A^{\mathrm{NE}}({\bf p^2}) \nn\\
B_{ij}^{\mathrm{NE}} &=& \left(\delta_{ij}-\frac{{\bf p_i}{\bf p_j}}{{\bf p^2}}\right) B_{\bot}^{\mathrm{NE}} ({\bf p^2})+ \frac{{\bf p_i}{\bf p_j}}{{\bf p^2}} B_{||}^{\mathrm{NE}} ({\bf p^2})
\eea

Hence, $A^{\mathrm{NE}}$, $B_{\bot}^{\mathrm{NE}}$ and $B_{||}^{\mathrm{NE}} $ can be written as:

\bea
A^{\mathrm{NE}} &=& A_i^{\mathrm{NE}} \frac{ {\bf p_i} }{{\bf p^2}} \nn\\
B_{\bot}^{\mathrm{NE}} &=& \frac{1}{2} B_{ij}^{\mathrm{NE}}  \left(\delta_{ij}-\frac{{\bf p_i}{\bf p_j}}{{\bf p^2}}\right) \nn\\
B_{||}^{\mathrm{NE}} &=& B_{ij}^{\mathrm{NE}}  \frac{{\bf p_i}{\bf p_j}}{{\bf p^2}} 
\eea

Utilising the notation used in Eq. \ref{abshort}, we express $A^{\mathrm{NE}}$, $B_{\bot}^{\mathrm{NE}}$ and $B_{||}^{\mathrm{NE}} $ in the following way:

\bea
A^{\mathrm{NE}} &=& \hat{\Phi} \left[ \left(1-\frac{{\bf p.p'}}{\bf p^2} \right) ~ \mathcal{R}^1_{\bf p,~q}\right]\nn\\
B_{\bot}^{\mathrm{NE}} &=&  \hat{\Phi} \left[ \frac{1}{4} \left({\bf p'}^2-\frac{({\bf p.p'})^2}{\bf p^2} \right) ~\mathcal{R}^2_{\bf p,~q}\right]\nn\\
B_{||}^{\mathrm{NE}} &=&  \hat{\Phi} \left[ \frac{1}{2} \left(\frac{({\bf p.p'})^2}{\bf p^2}-2{\bf p.p'}+{\bf p}^2 \right) ~ \mathcal{R}^2_{\bf p,~q}\right]
\label{abperpbpara}
\eea

We evaluate the transport coefficients in Eq. \ref{abperpbpara} using the standard techniques \cite{svetitsky}.

 \medskip
\begin{figure}[t]
\includegraphics[width=2.7in]{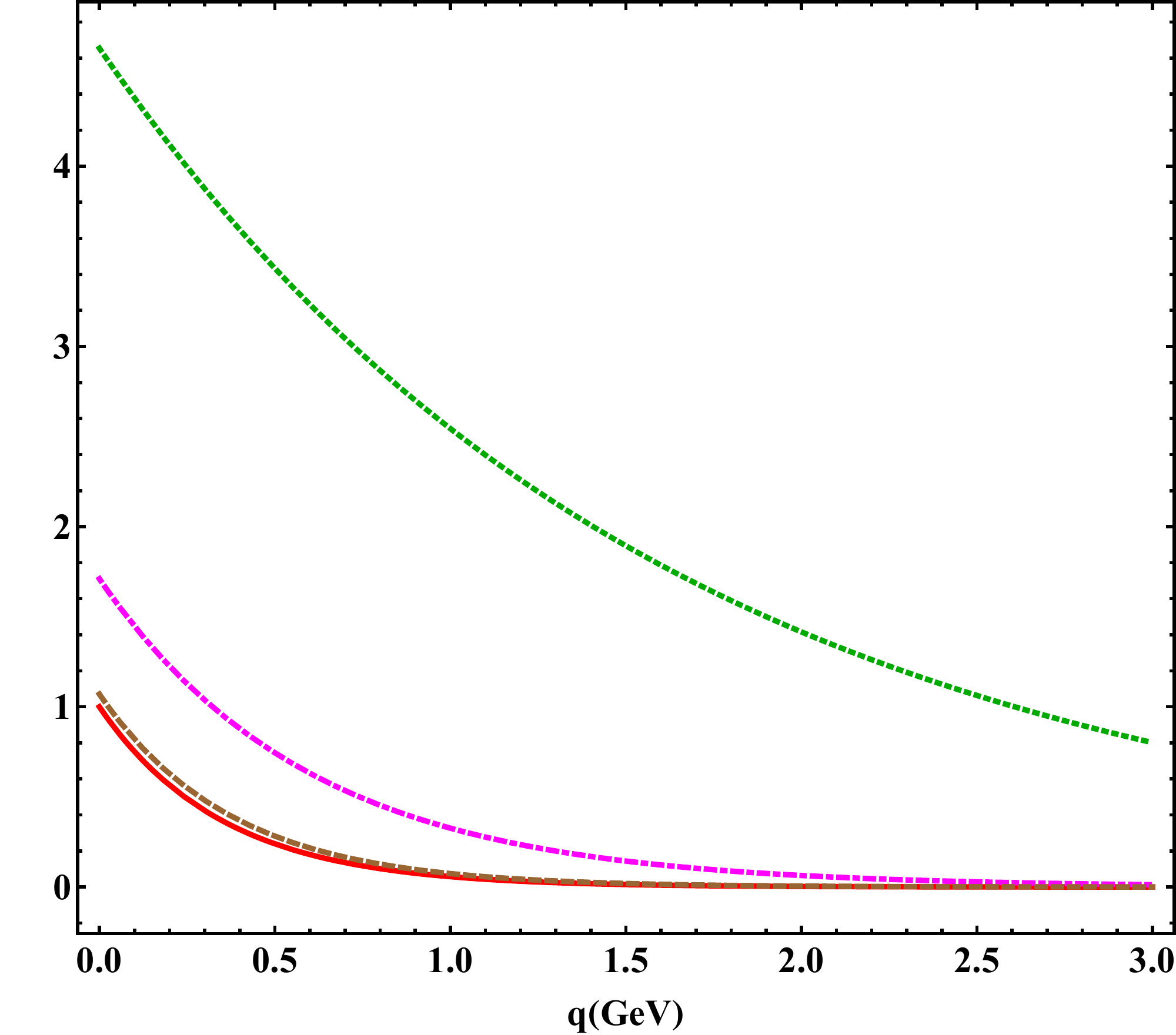}
\caption{Variation of $\mathcal{R}^1_{\bf p,q}$ and $f({\bf q}) $ (red/solid) with ${\bf q}$ (momenta of the medium particles). $\mathcal{R}^1_{\bf p,q}$ gradually approaches $f({\bf q})$ with decreasing $\delta q$ values: i) $\delta q=0.05$ (green/dotted), ii) $\delta q=0.01$ (magenta/dotdashed), iii) $\delta q=0.001$ (brown/dashed)}
\label{r1pq}
\end{figure}

\begin{figure}[h]
\includegraphics[width=2.7in]{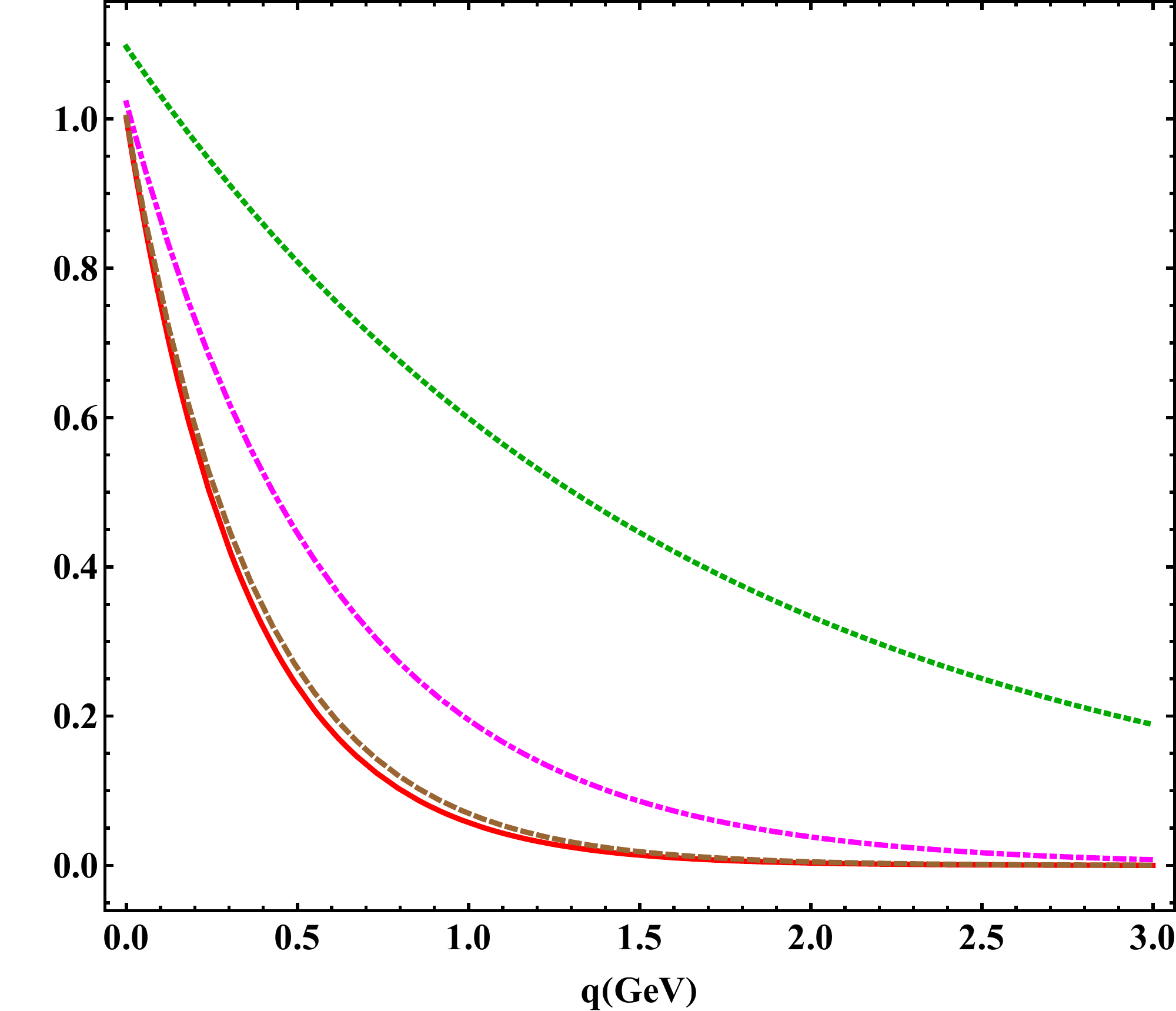}
\caption{Variation of $\mathcal{R}^2_{\bf p,q}$ and $f({\bf q}) $ (red/solid) with ${\bf q}$ (momenta of the medium particles). $\mathcal{R}^2_{\bf p,q}$ gradually approaches $f({\bf q})$ with decreasing $\delta q$ values: i) $\delta q=0.05$ (green/dotted), ii) $\delta q=0.01$ (magenta/dotdashed), iii) $\delta q=0.001$ (brown/dashed)}
\label{r2pq}
\end{figure}
\vskip 0.3in

\begin{figure}[h]
\includegraphics[width=2.8in]{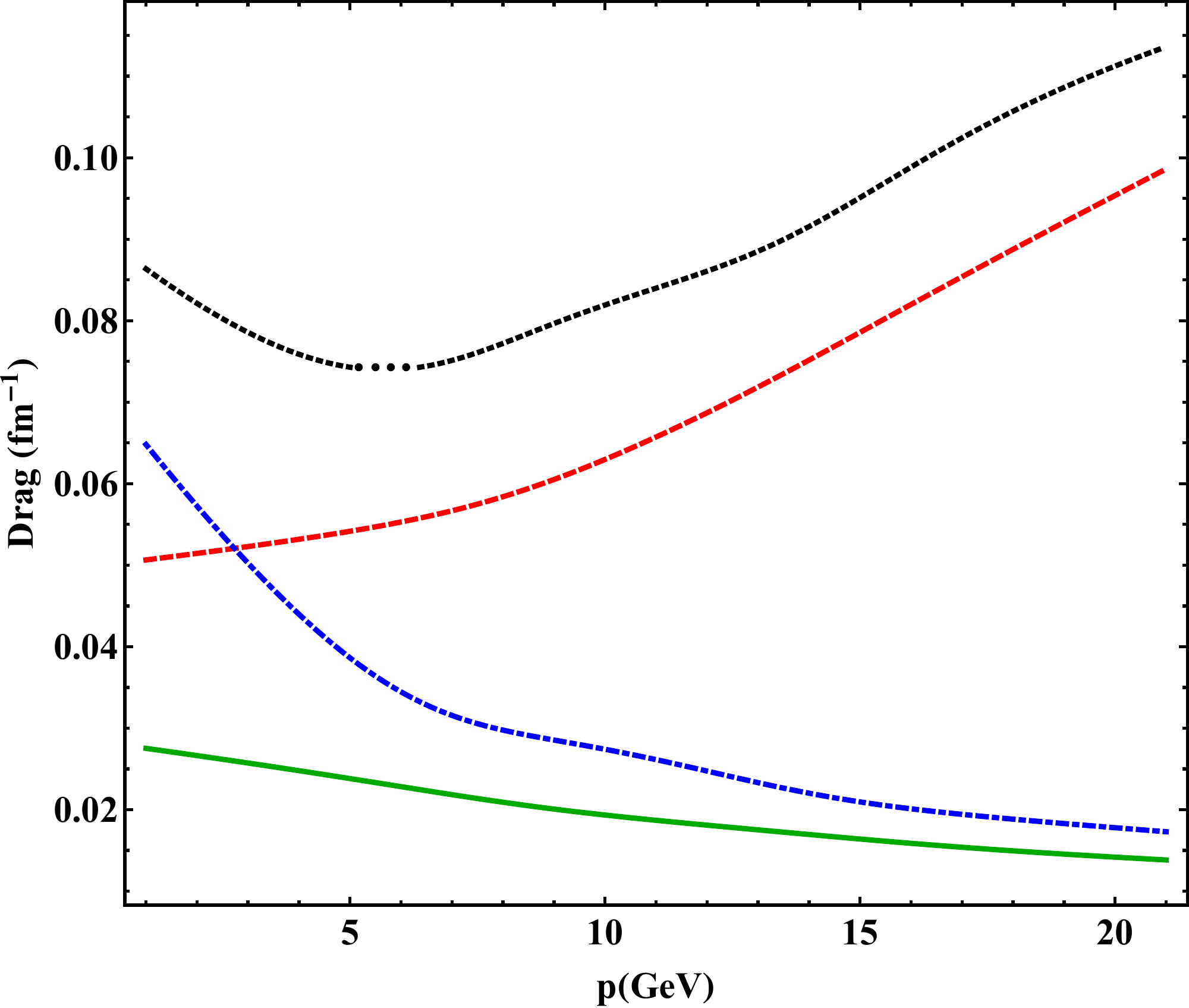}
\caption{Variation of the extensive and non extensive drag coefficients with momentum of the incoming heavy quark. The dotted (black) line represents the non-extensive drag for the charm quark and the dashed (red) line represents that for the bottom quark. The dot-dashed (blue) line and the solid (green) lines are the extensive drag coefficients for the charm and the bottom quark respectively.}
\label{dragptest}
\end{figure}

\medskip
\begin{figure}[h]
\includegraphics[width=2.8in]{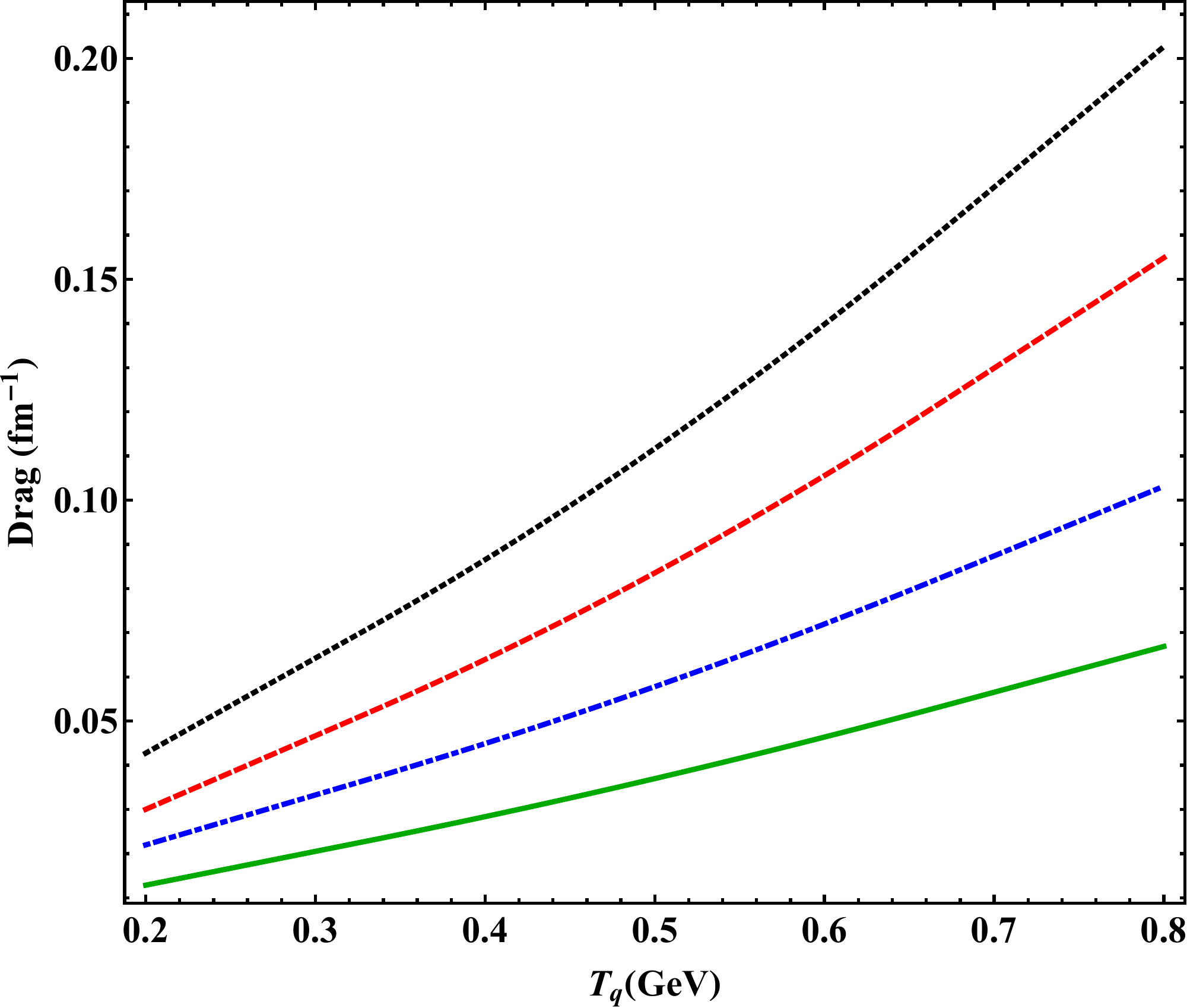}
\caption{Variation of the extensive and non extensive drag coefficients with temperature of the medium. The dotted (black) line represents the non-extensive drag for the charm quark and the dashed (red) line represents that for the bottom quark. The dot-dashed (blue) line and the solid (green) lines are the extensive drag coefficients for the charm and the bottom quark respectively.}
\label{dragtemp}
\end{figure}
\subsection{Non Linear Fokker Planck Transport Coefficients to Linear Fokker Planck Transport Coefficients}
\label{nlfptctolfptc}

Comparing Eqs. (\ref{nonexdragdiff}, \ref{exdragdiff}) it is clear that to get back the extensive Fokker Planck transport coefficients from the non 
extensive ones, the functions $\mathcal{R}^1_{\bf p,q}$ and $\mathcal{R}^2_{\bf p,q}$ should reduce to $f({\bf q})$ ($=\mathrm{Exp}(-{\bf q}/T)$, the momentum space medium particle distribution with medium temperature $T$) and we expect this reduction to take place in the limit $\delta q\rightarrow0$, {\it i.e.} when there exists no correlation between the probe and the medium particles. From Figs. (\ref{r1pq}, \ref{r2pq}), we can ascertain that with decreasing $\delta q$ values $\mathcal{R}^1_{\bf p,q}$ and $\mathcal{R}^2_{\bf p,q}$ both approach $f({\bf q})$. For $\delta q=0.001$ they are very close to $f({\bf q})$. Reducing the correlation further will result in the exact overlap. This proves that the modified expressions for the Fokker-Planck drag and diffusion coefficients give back the linear Fokker-Planck transport coefficients when there exists no correlation between the probe particle and the medium.

\section{Results and Discussion}
\label{results}

\begin{figure}[h]
\begin{center}
\includegraphics[width=2.8in]{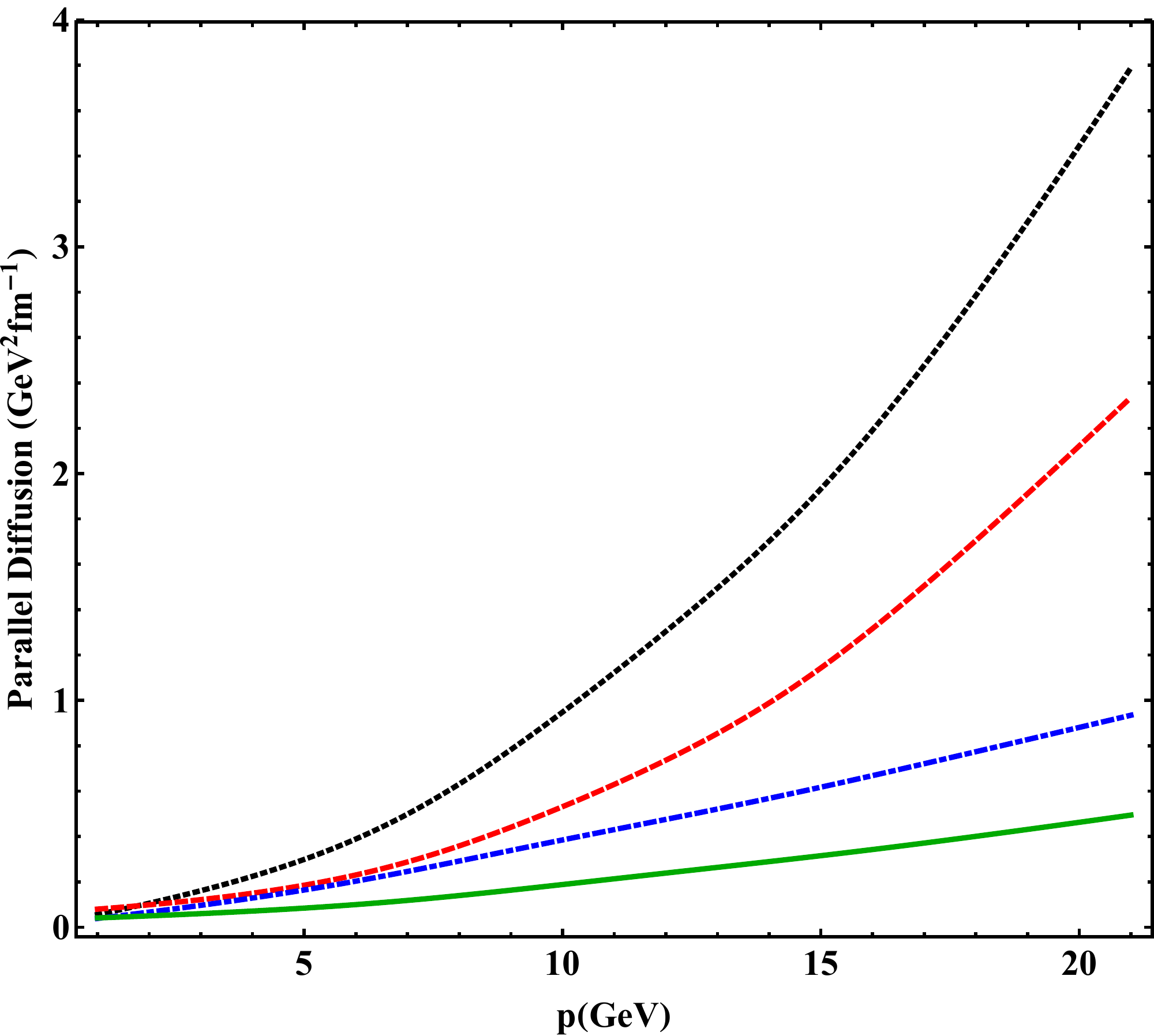}
\caption{Variation of the extensive and non extensive parallel diffusion coefficient with momentum of the incoming heavy quark. The dotted (black) line represents the non-extensive  for the charm quark and the dashed (red) line represents that for the bottom quark. The dot-dashed (blue) line and the solid (green) lines are the extensive drag coefficients for the charm and the bottom quark respectively.}
\label{diffparaptest}
\end{center}
\end{figure}

\begin{figure}[h]
\includegraphics[width=2.8in]{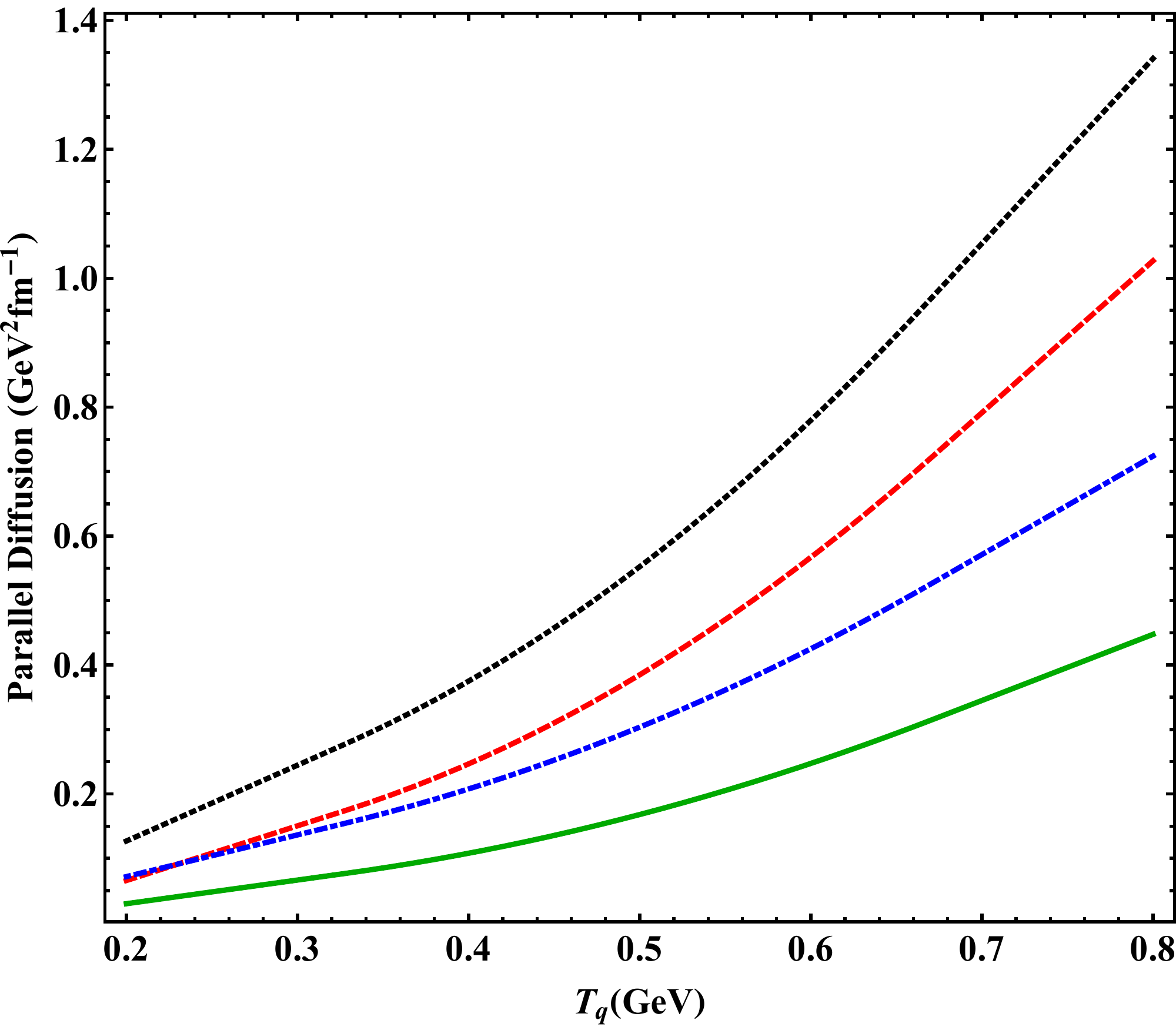}
\caption{Variation of the extensive and non extensive parallel diffusion coefficients with temperature of the medium. The dotted (black) line represents the non-extensive drag for the charm quark and the dashed (red) line represents that for the bottom quark. The dot-dashed (blue) line and the solid (green) lines are the extensive drag coefficients for the charm and the bottom quark respectively.}
\label{diffparatemp}
\end{figure}

\begin{figure}[t]
\begin{center}
\includegraphics[width=2.8in]{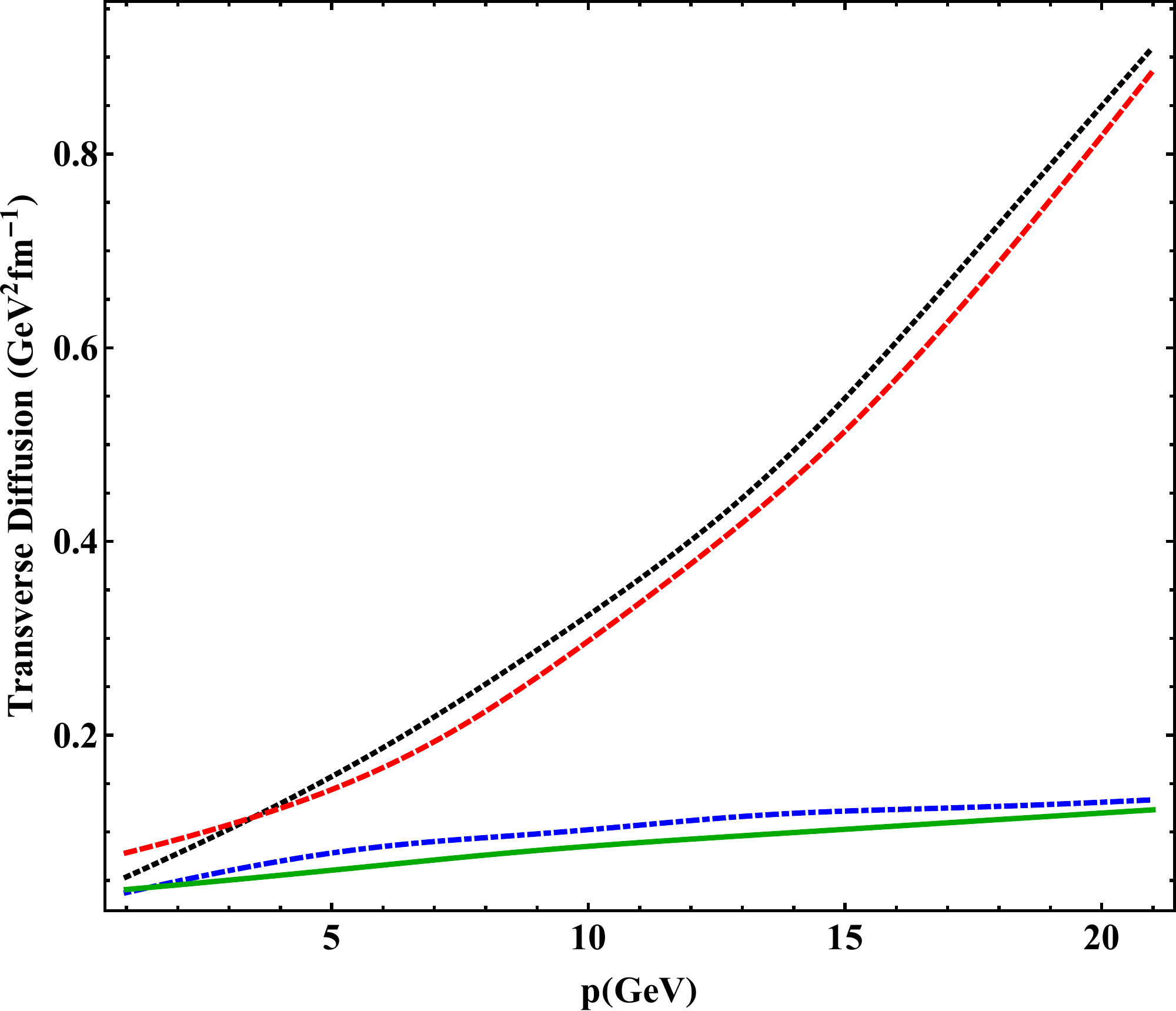}
\caption{Variation of the extensive and non extensive transverse diffusion coefficient with the momentum of the incoming heavy quark. The dotted (black) line represents the non-extensive transverse diffusion for the charm quark and the dashed (red) line represents that for the bottom quark. The dot-dashed (blue) line and the solid (green) lines are the extensive transverse diffusion coefficients for the charm and the bottom quark respectively.}
\label{diffperpptest}
\end{center}
\end{figure}

\begin{figure}[h]
\includegraphics[width=2.8in]{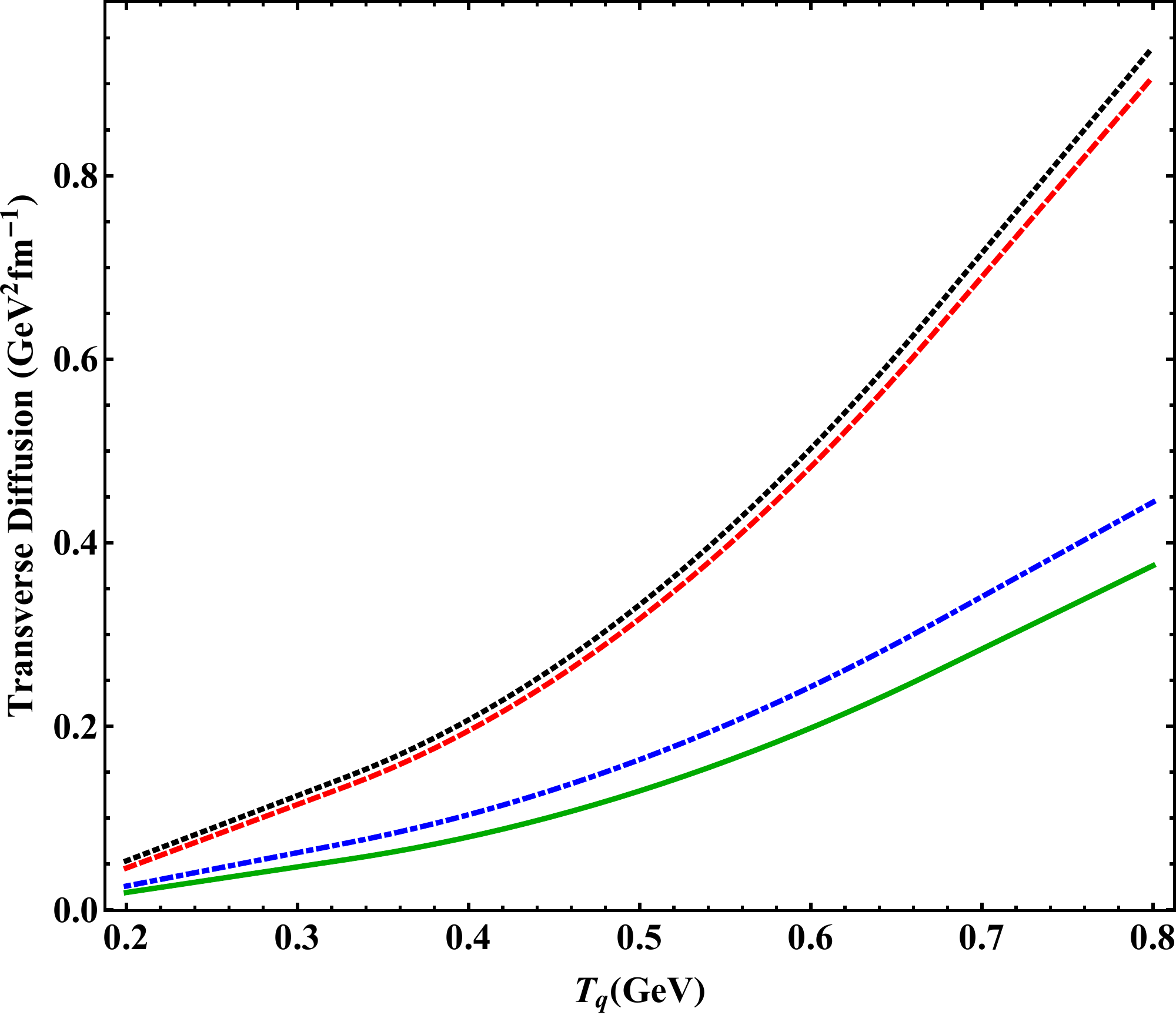}
\caption{Variation of the extensive and non extensive transverse diffusion coefficients with the temperature of the medium. The dotted (black) line represents the non-extensive transverse diffusion for the charm quark and the dashed (red) line represents that for the bottom quark. The dot-dashed (blue) line and the solid (green) lines are the extensive transverse diffusion coefficients for the charm and the bottom quark respectively.}
\label{diffperptemp}
\end{figure}

In our calculation we use 1.3 GeV and 4.2 GeV as the charm and bottom quark masses respectively. The parameter values characterizing the temperature profile in Eq. \eqref{tempprof} are \cite{tempevolution}: $T_p=290$ MeV, $a=5.99$ and $r_0=7.96$ fm. While the medium temperature $T_q$ is taken to be 350 MeV for the Figs. \ref{dragptest}, \ref{diffparaptest} and \ref{diffperpptest}, the incoming momenta are taken to be 5 GeV for Figs. \ref{dragtemp}, \ref{diffparatemp}, \ref{diffperptemp}. Also, while generating Figs. \ref{dragptest}-\ref{diffperptemp}, we put $\delta q=0.01$ as the correlation between the probe and the medium. In the calculations, we have considered only the heavy quarks elastically scattering with light quarks and gluons of the medium. The collisional matrix element has been taken from \cite{combridge}. The radiative Fokker-Planck drag and diffusion coefficients can be evaluated following the technique delineated in \cite{mbad}. This we reserve for our future work. In the plots of the Fokker-Planck transport coefficients we show both the (widely studied) extensive as well as the non-extensive cases for sake of comparison. The temperature variation of the collisional extensive transport coefficients can be compared with those from \cite{mbad} and they show more or less similar results. It is interesting to see that when we introduce a correlation between the heavy quark and the light quark/gluon in the medium at the same space time point, the transport coefficients increase. This can be attributed to the modification of the phase space due to correlation.

\medskip

There is a huge increase in the values of the non extensive transport coefficients. For charm or bottom quarks, travelling through a medium of 350 MeV temperature with 10 GeV momentum, the non extensive drag coefficient and the non extensive transverse diffusion coefficient are $\sim$ 3 times more than their extensive counterpart. The corresponding factor is $\sim$ 2.5 for the parallel diffusion coefficient. Also, the heavy quark non extensive drag increases with the momentum as opposed to the trend shown by its extensive counterpart.

\section{Summary, Conclusion and Outlook}
\label{summary}

To summarize, we have calculated the Fokker-Planck drag and diffusion coefficients of heavy quarks (charm and bottom) traversing through a medium of quarks and gluons and interacting elastically with them. The novelty of this work lies in the introduction of the correlation of the incoming heavy quarks with the medium particles. We observe that the transport coefficients are substantially modified when we introduce correlation. Also, in the vanishing correlation limit we get back the extensive transport coefficients.

In the present calculations, we have considered the collisional processes only. As already  mentioned, radiative scattering processes will also be important particularly in the high momentum region. Treatment of the radiative processes (heavy quark scattering with light quark to emit single gluon is one such example) can be treated following the techniques outlined in \cite{mbad} and we reserve the work for future. 

Combining drag and the stopping power $dE/dx$ (energy loss per unit time time divided by the particle speed) we can define a relativistically invariant quantity \cite{rafelskiwaltonprl} and hence the present calculation can directly lead to the calculation of the stopping power in the ambience of fluctuating temperature. Also, using the non extensive transport coefficients we can try to solve the non linear Fokker-Planck equation to find the evolution of the incoming heavy quark distribution. The ratio of the final distribution to the initial distribution can be compared with the experimentally observed nuclear suppression factor ($R_{\mathrm{AA}}$) of heavy quarks. The results can be compared/contrasted with the results obtained in the earlier works in this direction in the Refs. \cite{tsallisraaepja}, \cite{tsallisraaflow}.

\section*{Acknowledgement}
TB acknowledges the University Research Committee, University of Cape Town, South Africa for support.

\end{document}